\documentclass[12pt,preprint2]{aastex}

\begin{document}

\title{Discovery of Non-radial pulsations in PQ Andromedae}

\author{Karen M. Vanlandingham}
\affil{Columbia University, Department of Astronomy, New York, NY 10027}
\email{kmv14@columbia.edu}

\author{Greg J. Schwarz}
\affil{Steward Observatory, University of Arizona, Tucson, AZ 85721}
\email{gschwarz@as.arizona.edu}

\and

\author{Steve B. Howell}
\affil{WIYN Observatory and National Optical Astronomy Observatory, 950 North Cherry Avenue, Tucson, AZ 85726}
\email{howell@noao.edu}

\begin{abstract}
We have detected pulsations in time-series photometry of the
WZ Sge dwarf nova PQ And.  The strongest peak in the power spectrum
occurs at a period of 10.5 minutes.  Similar periods have been
observed in other WZ Sge systems and are attributed to 
ZZ Ceti type non-radial pulsations.  There is no indication in 
the photometry of an approximately 1.7 hour orbital period as 
reported in previous spectroscopic observations.
\end{abstract}


\section{Introduction}
PQ And is a dwarf nova with three known outbursts in the last $\sim$ 70 
years (Richter 1990).  The last outburst on 21 March, 1988 reached a 
visual magnitude of 10 (Hurst 1988).  It had many of the characteristics
of the WZ Sge class of dwarf novae including a large outburst 
amplitude
and strong Balmer emission superimposed over broad absorption at quiescence 
(Howell et al. 1995).  However, to be in the WZ Sge class, PQ And must have
an orbital period below the period gap.  Time series optical spectra of
PQ And were obtained by Schwarz et al. (2004) to determine the orbital
period of the system.  They found a period of 1.7 hours from the fit to 
the radial velocity curve as determined from the convolution of double 
Gaussians to the line profiles (Schafter 1985).  Unfortunately the
derived period was similar in length to the total observing time and 
couldn't be classified as a definite period due to undersampling.  
Nevertheless the H$\alpha$ emission showed a clear periodic change in 
its blue and red components during the observing run which supported 
a short orbital period typical of WZ Sge systems.  An optically thin disk,
showing the WD absorption, is indicative of a short period, low-$\dot{m}$
system (i.e. below the period gap).  Likewise, the lack 
of an accretion disk in the optical spectrum placed a firm upper limit 
of 3 hours on the orbital period of PQ And.

Schwarz et al. also determined the effective temperature and 
surface gravity of the white dwarf (WD) using synthetic spectra 
from model atmospheres.  The best fits gave T$_{eff}$ = 12,000 $\pm$ 1,000 K 
and log($g$) = 7.7 $\pm$ 0.3 (cgs units) which placed PQ And in the 
region of the ZZ Ceti instability strip (Bergeron et al. 1995; 2004).  They 
noted that with its low accretion rate, PQ And was an excellent candidate 
to search for non-radial oscillations which have recently been observed 
in other WZ Sge novae.

In this paper we present the results of a photometric campaign to search 
for non-radial pulsations in PQ And.  Section 2 provides details of the 
observations.  The analysis of the data are given in Section 3 and our
conclusions follow in Section 4.

\section{Observations}

Our photometric observations were carried out using the Orthogonal 
Parallel Transfer Imaging Camera (OPTIC, see Howell et al, 2003) 
at the WIYN observatory 3.5-m telescope located on Kitt Peak.
OPTIC is the only prototype orthogonal transfer CCD
imager operating (see Tonry et al., 2002) and 
consists of two 2K by 4K CCID-28 OTCCDs in a single dewar
mounted adjacent to each other with a small gap in between the chips.
OPTIC is controlled by standard SDSU-2 electronics running custom microcode
and reads out the two OTCCDs via 4
video channels, one located in each corner of the device.
OPTIC has a read noise of $<$4 electrons when
read at a normal rate of 160 kpix/sec and a gain of 1.45 e/ADU.

We used OPTIC in conventional mode placing the target star and all comparison 
stars of interest in one of the CCDs. 
Two time series data sets were obtained, the first was on the night of 14
September 2004 UT and the second on 13 October 2004 UT. The September
observations consisted of $\sim$125, 45 second observations using a 
Johnson V filter.
The CCD was binned 2 X 2 (the seeing on this night was 1.6") and the readout
time was 8 seconds. The October observations consisted of $\sim$100 60 second 
V-band integrations using 1 X 1 binning (readout time was 24 seconds) and 
the seeing was 0.5".

The data were reduced using the standard IRAF packages.  Relative photometry
was performed using four different background stars as references.  Figure 1
shows a plot of the magnitude difference as a function of time for both nights.
The squares represent the difference between two of the reference stars while
the triangles represent the difference between PQ And and one of the reference
stars.  The horizontal axis is the time since the start of observations for
each night.
The light curves of both nights clearly show fluctuations of up to
$\sim$0.1 magnitudes. 
All reference stars and PQ And were of similar magnitude therefore
the scatter in the reference star data gives an estimate of the errors in
the magnitude change for PQ And.  From Figure 1 we see that the magnitude
errors are on order of 0.015.

\begin{figure}
\plotone{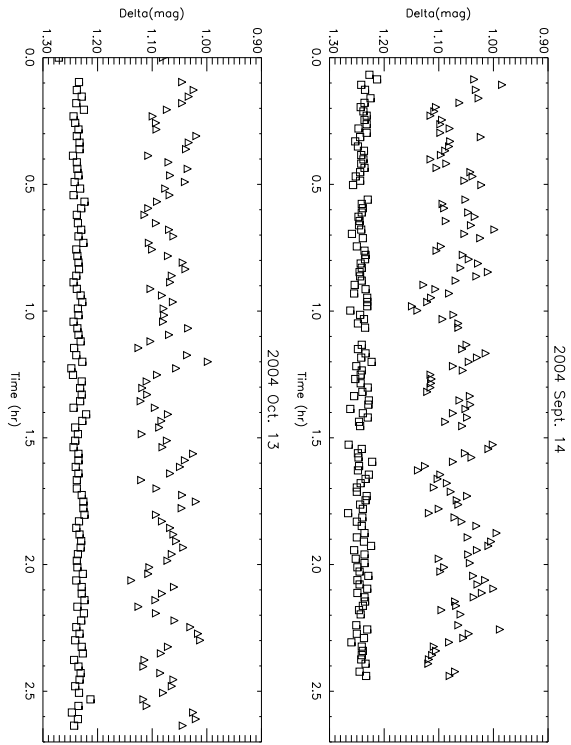}
\caption{The magnitude difference between two reference stars (squares) and
PQ And and one of the reference stars (triangles) for each night.  t$_0$ is
taken as the time of the first observation of each night.}
\end{figure}

\section{Analysis}

To search for any periodicities in the light curves the data were put 
through a Fourier transform routine.  The resulting power spectra for each
date are shown in Figure 2.  Based on the length of time PQ And
was observed and the sampling rate of each night, we searched the frequency 
space from 0.5 to 30 hr$^{-1}$.
To determine the level of significance for peaks in the power spectrum we
added a tracer signal to our data and ran it through the Fourier routine
until we were unable to detect it.  From this we found that peaks with a
power less than $\sim10^{-5}$ were not significant.
Three strong peaks appear in the 
power spectra of both nights.  The strongest peak is found at a frequency 
of 5.6 hr$^{-1}$, a period of 10.5 minutes, in both spectra.  We also see 
what appear to be harmonics of this peak at frequencies of $\sim$2.8 hr$^{-1}$ 
and $\sim$1.4 hr$^{-1}$.  At higher frequencies the two power spectra differ
slightly.  In the September data we see no significant peaks beyond the
three already mentioned while the October power spectrum shows a few
small peaks at higher frequencies.  These peaks are found at 7.8, 8.4, 9.2,
22.5, and 29 hr$^{-1}$ (periods in the range of 2-8 minutes).  The difference
between the two nights could be due to many things including sampling rates,
seeing effects, and variability in the WD itself.

\begin{figure}
\plotone{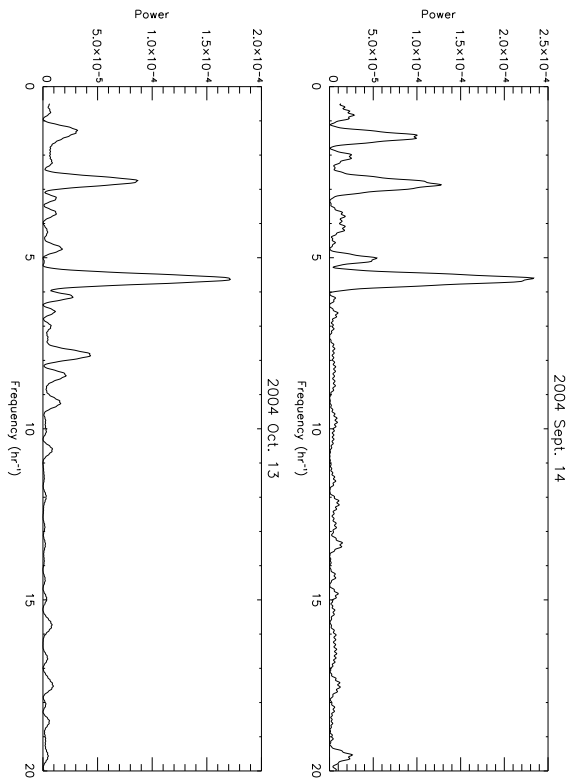}
\caption{The individual power spectra of the two dates.  Both dates show
peaks near frequencies of 5.6, 2.8 and 1.5 hr$^{-1}$.  The observations 
from October show a few additional minor peaks at longer frequencies.}
\end{figure}

Figure 3 shows the data phased to a period of 10.5 minutes and put in phase
bins.  The error bars have been determined
from the scatter in the light curve of the reference stars.  The data 
show an overall change in magnitude of about 0.1.

\begin{figure}
\plotone{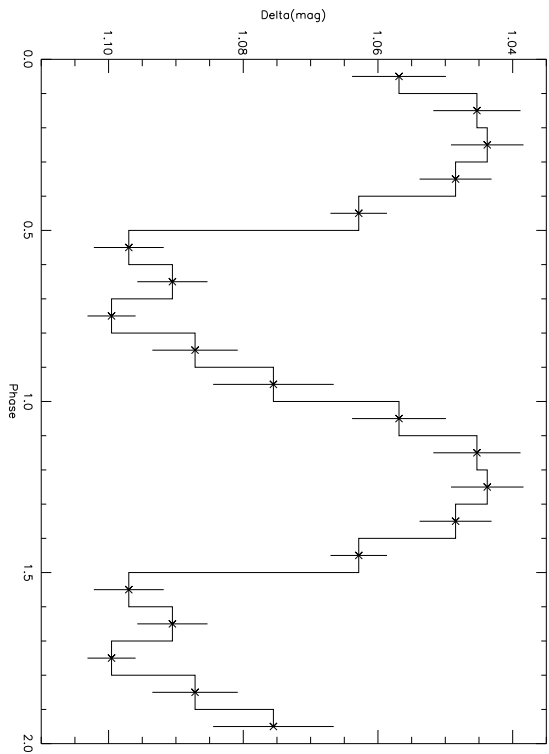}
\caption{The data from both nights phased to a period of 10.5 minutes and
binned.}
\end{figure}

\section{Discussion}
The most significant period we find, 10.5 minutes, is much shorter than 
any expected orbital period for a CV. Several WZ Sge systems have recently 
been found with similar short period, low amplitude oscillations
(Araujo-Betancor et al 2005; Townsley, Arras \& Bildsten 2004;
Woudt \& Warner 2004; van Zyl et al 2004).  These periodic fluctuations
have been attributed to ZZ Ceti like non-radial pulsations of the 
WD primary.  In the five systems reported to date, there appears
to be a pattern in the periods found.  All of the stars show periods 
around 10 minutes and then several other periods on shorter time scales 
(3-6 minutes).  Long term monitoring shows that these periods are variable 
between subsequent observations.
Our findings for PQ And match this general pattern.  Some of
the other ZZ Ceti systems in the literature have more complicated power spectra,
showing peaks at longer frequencies that are multiples or combinations of 
the shorter frequency peaks.  We don't see these patterns in our data for PQ 
And however
it could be that we have insufficient data to detect these complexities.
Most of the other ZZ Ceti systems have far more extensive data sets.

The radial velocity analysis of Schwarz et al. (2004) implied an orbital 
period of $\sim$102 minutes for PQ And.  Figure 4 shows the power spectrum 
of our data in the frequency range of 0.5 to 2 hr$^{-1}$ (120 to 30 minutes).  
The strongest peak is the 2nd harmonic of the 10.5 minute pulsation
period.  There is 
a smaller peak structure at about 0.7 hr$^{-1}$ (86 minutes) but it 
doesn't fit the previously derived orbital period and it is very close to
our significance threshold.  
It is disconcerting that the orbital period as determined from the 
radial velocities is not seen in the power spectrum of the photometry.  
One explanation is that the PQ And 
system is at a low inclination.
If this is the case then orbital motions would cause the observed
periodicity in the H$\alpha$ red and blue emission components but no
photometric eclipses would be seen.  
We can place an upper limit on the system inclination
using the estimated mass ratio from Schwarz et al. (2004) and the minimum
inclination for primary eclipses (see equations 2.5c and 2.64 from 
Warner 1995).  A $q\sim0.3$ places an upper inclination limit of about 
74$^\circ$ for PQ And.

\begin{figure}
\plotone{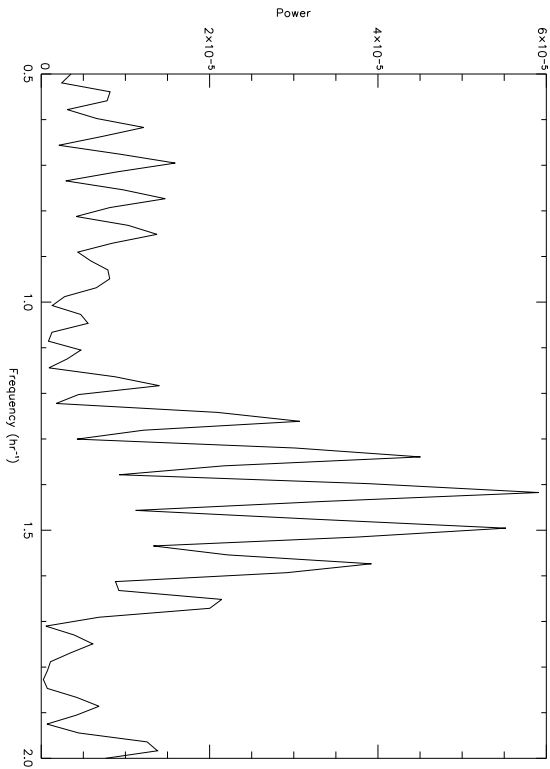}
\caption{The power spectrum of the combined data from 0.5 to 2 hr$^{-1}$.
There is no orbital period peak at 0.59 hr$^{-1}$ as suggested by the 
radial velocity analysis of Schwarz et al. (2004).}
\end{figure}

The available evidence points to PQ And containing a ZZ Ceti white dwarf.
The primary pulsation period we find is 10.5 minutes which is similar to 
those found in other WZ Sge type systems with ZZ Ceti like white dwarf
parameters and pulsations.  More extensive observations may reveal 
alternate pulsation modes as has been found in other ZZ Ceti systems.
We find no evidence for the 1.7 hour orbital period reported by 
Schwarz et al (2004) in our data.

\acknowledgments
We would like to thank Gary Schmidt for use of his Fourier analysis programs.
We would also like to thank the anonymous referee for their many useful
comments which helped to improve this paper.

\end{document}